\documentclass[a4paper]{article}
\usepackage{arxiv}
\pdfoutput=1
\newif\ifDoubleBlind
\DoubleBlindfalse

\usepackage[T1]{fontenc}
\usepackage[utf8]{inputenc}

\usepackage{booktabs}
\usepackage{makecell}
\usepackage{relsize}
\usepackage{amssymb}
\usepackage{siunitx}
\usepackage[labelformat=parens]{subcaption}

\usepackage[normalem]{ulem}

\usepackage{siunitx}

\newcommand{\astr}[1]{\renewcommand{\arraystretch}{#1}}
\newcommand{\ie}{i.e.}
\newcommand{\eg}{e.g.}
\newcommand{\ea}{et\ al.}
\newcommand{\bigO}{\mathcal{O}}
\newcommand{\smallO}{o}
\newcommand{\softO}{\tilde{\mathcal{O}}}

\newcommand{\Kilo}{\ensuremath{\mathrm{k}}}
\newcommand{\Mega}{\ensuremath{\mathrm{m}}}
\newcommand{\AlgLongName}[1]{\textit{#1}}

\newcommand{\VertexSet}{V}

\newcommand{\initE}{m}
\newcommand{\avgE}{\overline{m}}
\newcommand{\avgD}{\overline{d}}
\newcommand{\finE}{M}
\newcommand{\Arc}[2]{(#1, #2)}
\newcommand{\Degree}[1]{\ensuremath{\mathrm{deg}(#1)}}
\newcommand{\InDegree}[1]{\ensuremath{\mathrm{deg}^\mathsmaller{-}(#1)}}
\newcommand{\OutDegree}[1]{\ensuremath{\mathrm{deg}^\mathsmaller{+}(#1)}}

\newcommand{\Source}{s}

\newcommand{\Ratio}{\rho}
\newcommand{\Card}[1]{\ensuremath{|#1|}}
\newcommand{\List}[1]{\ensuremath{\mathcal{L}}}
\newcommand{\InNeighborList}[1]{\ensuremath{\mathcal{N}^-}}
\newcommand{\OutNeighborList}[1]{\ensuremath{\mathcal{N}^+}}
\newcommand{\InEdgesList}[1]{\ensuremath{\mathcal{E}^-}}
\newcommand{\OutEdgesList}[1]{\ensuremath{\mathcal{E}^+}}
\newcommand{\TreeEdgeIndex}[1]{\Code{e}[#1]}
\newcommand{\Level}[1]{\Code{l}[#1]}
\newcommand{\Queue}[1]{\Code{Q}}
\newcommand{\Set}[1]{\left\{#1\right\}}
\newcommand{\Range}[1]{\left[#1\right]}
\newcommand{\Ops}{\sigma}

\newcommand{\Updates}{\delta}
\newcommand{\AddsRatio}{\Updates{}_{\mathsmaller{+}}}

\newcommand{\SSR}{\textsl{SSR}}
\newcommand{\APRlong}{all-pairs reachability}

\newcommand{\SDFS}{\texttt{SDFS}}
\newcommand{\SBFS}{\texttt{SBFS}}
\newcommand{\CDFS}{\texttt{CDFS}}
\newcommand{\CBFS}{\texttt{CBFS}}
\newcommand{\LDFS}{\texttt{LDFS}}
\newcommand{\LBFS}{\texttt{LBFS}}
\newcommand{\SIM}{\texttt{SI}}
\newcommand{\SIMp}[1]{\texttt{SI(#1)}}
\newcommand{\ES}{\texttt{ES}}
\newcommand{\ESp}[1]{\texttt{ES(#1)}}
\newcommand{\SES}{\texttt{SES}}
\newcommand{\SESp}[1]{\texttt{SES(#1)}}
\newcommand{\MES}{\texttt{MES}}
\newcommand{\MESp}[1]{\texttt{MES(#1)}}
\newcommand{\FlagR}{\ensuremath{\mathtt{R}}}
\newcommand{\FlagNR}{\ensuremath{\overline{\mathtt{R}}}}
\newcommand{\FlagRQ}{\ensuremath{\mathtt{R}?}}
\newcommand{\FlagSF}{\ensuremath{\mathtt{SF}}}
\newcommand{\FlagNSF}{\ensuremath{\overline{\mathtt{SF}}}}
\newcommand{\FlagSFQ}{\ensuremath{\mathtt{SF}?}}
\newcommand{\RestartLimit}{\ensuremath{\beta}}
\newcommand{\RestartRatio}{\ensuremath{\Ratio}}

\newcommand{\Code}[1]{\texttt{#1}}
\newcommand{\Reachable}{\Code{reachable}}
\newcommand{\Unreachable}{\Code{unreachable}}
\newcommand{\Unknown}{\Code{unknown}}
\newcommand{\Routine}[1]{\textsf{#1}}

\newcommand{\Tool}[1]{\textsf{#1}}
\newcommand{\KONECT}{\Tool{KONECT}}
\newcommand{\SNAP}{\Tool{SNAP}}
\newcommand{\ALGORA}{\Tool{Algora}}

\newcommand{\TabLabel}[1]{\label{tab:#1}}
\newcommand{\Table}[1]{Table~\ref{tab:#1}}

\newcommand{\Figure}[1]{Figure~\ref{fig:#1}}
\newcommand{\Figures}[2]{Figures~\ref{fig:#1} and~\ref{fig:#2}}

\newcommand{\FigureSubRange}[2]{Figures~\ref{fig:#1}--\subref{fig:#2}}
\newcommand{\SectLabel}[1]{\label{sect:#1}}
\newcommand{\Section}[1]{Section~\ref{sect:#1}}

\usepackage{color}

\usepackage{tikz}
\usetikzlibrary{matrix}
\usepackage{pgfplots}
\pgfplotsset{compat=1.14}

\title{Fully Dynamic Single-Source Reachability in Practice: An Experimental Study}
\author{Kathrin Hanauer}{University of Vienna, Faculty of Computer Science, Vienna, Austria}{kathrin.hanauer@univie.ac.at}{https://orcid.org/0000-0002-5945-837X}{}
\author{Monika Henzinger}{University of Vienna, Faculty of Computer Science, Vienna, Austria}{monika.henzinger@univie.ac.at}{https://orcid.org/0000-0002-5008-6530}{}
\author{Christian Schulz}{University of Vienna, Faculty of Computer Science, Vienna, Austria}{christian.schulz@univie.ac.at}{https://orcid.org/0000-0002-2823-3506}{}
\authorrunning{K.\ Hanauer, M.\ Henzinger, C.\ Schulz}
\Copyright{Kathrin Hanauer, Monika Henzinger, Christian Schulz}

\funding{The research leading to these results has received funding
from the European Research Council under the European Union’s
Seventh Framework Programme (FP/2007-2013) / ERC Grant
Agreement no.\ 340506.}%

\date{}

\begin{document}
\maketitle
\begin{abstract}%
Given a directed graph and a source vertex, the \emph{fully dynamic single-source
reachability} problem is to maintain the set of vertices that are reachable from
the given vertex, subject to edge deletions and insertions.
It is one of the most fundamental problems on graphs and appears directly or
indirectly in many and varied applications.
While there has been theoretical work on this problem, showing both linear
conditional lower bounds for the fully dynamic problem and insertions-only and
deletions-only upper bounds beating these conditional lower bounds, there has
been no experimental study that compares the performance of fully dynamic
reachability algorithms in practice.
Previous experimental studies in this area concentrated only on the more
general all-pairs reachability or transitive closure problem and did not use
real-world dynamic graphs.

In this paper, we bridge this gap by empirically studying an extensive set of
algorithms for the single-source reachability problem in the fully dynamic
setting.
In particular, we design several fully dynamic variants of well-known
approaches to obtain and maintain reachability information with respect to a
distinguished source.
Moreover, we extend the existing insertions-only or deletions-only upper bounds
into fully dynamic algorithms.
Even though the worst-case time per operation of all the fully dynamic
algorithms we evaluate is at least linear in the number of edges in the graph
(as is to be expected given the conditional lower bounds) we show in our
extensive experimental evaluation that their performance differs greatly, both
on generated as well as on real-world instances.
\par %
\end{abstract}

\section{Introduction}%
Many real-world problems can be expressed using graphs and in turn be solved
using graph algorithms.
Often, the underlying graphs or input instances change over time, \ie, vertices
or edges are inserted or deleted as time is passing.
In a social network, for example, users sign up or leave, and relations between
them may be created or removed over time.
Another typical example is the OpenStreetMap road network, which is permanently
subject to change as roads are built or (temporarily) closed, or simply because
new information is added to the system by users.
Given a concrete graph problem, computing a new solution for every change that
occurs in the graph can be an expensive task on huge networks or where hardware
resources are scarce, and ignores the previously gathered information on the
instance under consideration.
Hence, a whole body of algorithms and data structures for dynamic graphs has
been discovered in the last decades.
It is not surprising that dynamic algorithms and data structures are in most
cases more difficult to design and analyze than their static counterparts.

Typically, dynamic graph problems are classified by the types of updates
allowed.
A problem is said to be fully dynamic if the update operations include
insertions \emph{and} deletions of edges.
If only insertions are allowed, the problem is called incremental; if only
deletions are allowed, it is called decremental.

One of the most basic questions that one can pose is that of reachability in
graphs, \ie, answering the question whether there is a directed path between
two distinct vertices.
Already this simple problem has many applications such as in program
analysis~\cite{reps1998program}, spanning from compiler optimization to
software security, or in the analysis of social or hyperlink networks---eg,
whether somebody is a friend of a friend, relationship detection, or
centrality measures.
It also appears in computational biology, when analyzing metabolic or
protein-protein interaction networks~\cite{10.1093/nar/gkq1207}.
Additionally, it is a very important subproblem in a wide range of more
complex (dynamic) algorithms such as in the computation of (dynamic) maximum
flows~\cite{fordfulkerson1956,Edmonds1972,goldberg2011maximum}, which
in turn have manifold applications.
However, state-of-the-art implementations typically run (slow) static
breadth-first searches repeatedly to accomplish this task since there is no
knowledge about the performance of more sophisticated algorithms in practice.

The single-source reachability problem has been extensively analyzed
theoretically.
The \emph{fully dynamic single-source reachability} (\SSR{}) problem is to
maintain the set of vertices that are reachable from a given \emph{source
vertex}, subject to edge deletions and insertions.
For the static version of the problem, \ie, when the graph does not change over
time, reachability queries can be answered in constant time after linear
preprocessing time by running, \eg{}, breadth-first search from the source
vertex and marking each reachable vertex.
This approach can be extended in the insertions-only case by using incremental
breadth-first search so that each insertion takes \emph{amortized} constant time
and each query takes constant time.
In the fully dynamic case, however, conditional lower bounds~\cite{HKNS15,AW14}
give a strong indication that no faster solution than the naive recomputation
from scratch is possible after each change in the graph.
There has been a large body of research on the deletions-only
case~\cite{ES81,HKN14,CHILP16}, leading to a $\bigO(\log^4{n})$~\cite{BPW19}
amortized expected time per deletion.
However, to the best of our knowledge, there has been no prior experimental
evaluation of fully dynamic \emph{single-source} reachability algorithms.

In this paper, we attempt to start bridging this gap by empirically studying an
extensive set of algorithms for the single-source reachability problem in the
fully dynamic setting.
In particular, we design several fully dynamic variants of well-known static
approaches to obtain and maintain reachability information with respect to a
distinguished source.
Moreover, we modify existing algorithms that provide theoretical guarantees
under the insertions-only or deletions-only setting to be fully dynamic.
We then perform an extensive experimental evaluation on random as well as
real-world instances in order to compare the performance of these algorithms.
In addition, we introduce and assess different thresholds that trigger
a recomputation from scratch to mitigate extreme update costs, which
turned out to be very effective.
Our results further show that making the insertions-only or deletions-only
algorithms fully dynamic leads to faster algorithms than ``dynamizing'' static
breadth-first or depth-first search.
\section{Preliminaries}%
\label{prelim}
\subsection{Basic Concepts}
Let $G = (V, E)$ be a directed graph with vertex set $V$ and edge set
$E$.
Throughout this paper, let $n=|V|$ and $m=|E|$.
The \emph{density} of $G$ is $d = \frac{m}{n}$.
An edge $\Arc{u}{v} \in E$ has \emph{tail} $u$ and \emph{head} $v$
and $u$ and $v$ are said to be \emph{adjacent}.
$\Arc{u}{v}$ is said to be an \emph{outgoing} edge or \emph{out-edge} of $u$
and an \emph{incoming} edge or \emph{in-edge} of $v$.
The \emph{outdegree} $\OutDegree{v}$/\emph{indegree}
$\InDegree{v}$/\emph{degree} $\Degree{v}$ of a vertex $v$ is its number of
(out-/in-) edges.
A sequence of vertices $s \to \cdots \to t$ such that each pair of consecutive
vertices is connected by an edge, is called an \emph{$s$-$t$ path} and $s$ can
\emph{reach} $t$.

\begin{table*}[tb]
\caption{Algorithms and abbreviations overview.}
\TabLabel{algorithms-overview}
\centering
\astr{1.2}
\begin{tabular}{@{}ll@{\hskip 33pt}ll@{}}
\toprule
Algorithm & Long name & Algorithm & Long name \\
\midrule
\SDFS{} /
\CDFS{} /
\LDFS{}
& Static/Caching/Lazy DFS
& \ESp{\RestartLimit/\RestartRatio}
& Even-Shiloach
\\
\SBFS{} /
\CBFS{} /
\LBFS{}
& Static/Caching/Lazy BFS
& \MESp{\RestartLimit/\RestartRatio}
& Multi-Level Even-Shiloach
\\
\SIMp{\FlagRQ{}/\FlagSFQ{}/\RestartRatio}
& Simple Incremental
& \SESp{\RestartLimit/\RestartRatio}
& Simplified Even-Shiloach
\\
\bottomrule
\end{tabular}
\end{table*}

A \emph{dynamic graph} is a directed graph $G$ along with an ordered sequence
of updates, which consist of edge insertions and deletions.

The paper deals with the \emph{fully dynamic single-source reachability
problem}~(\SSR{}):
Given a directed graph and a source vertex~$s$, answer reachability queries
starting at $s$, subject to edge insertions and~deletions.
\subsection{Related Work}%
In an incremental setting, where edges may only be inserted, but never are deleted,
a total update time of $\bigO(m)$ for $m$ insertions can be achieved by an
incremental breadth-first or depth-first search starting from the source vertex.
For a long time, the best algorithm to handle a series of $m$ edge
deletions and no insertions required a total update time of $\bigO(mn)$ and
actually solved the more general all-pairs shortest path problem.
The algorithm is due to Even and Shiloach~\cite{ES81,HK95,King99} and
maintains a breadth-first tree under edge deletions.
It is widely known as \emph{ES tree}.
Recently, Henzinger \ea{}~\cite{HKN14,HKN15} broke the $\bigO(mn)$ time
barrier by giving a probabilistic algorithm with an expected total update time
of $\bigO(mn^{0.9+\smallO(1)})$.
Shortly thereafter, Chechik~\ea{}~\cite{CHILP16} improved this result further
by presenting a randomized algorithm with $\softO(m\sqrt{n})$ total update
time.
Only lately, Bernstein~\ea{}~\cite{BPW19} showed that reachability information
in the decremental setting can be maintained in $\bigO(m\log^4 n)$ total
expected update time.
Whereas these algorithms all operate on general graphs,
Italiano~\cite{Italiano88} observed that a running time of $\bigO(m)$ may
indeed be achieved also in the decremental setting if the input graph is
acyclic.
Finally, if both edge insertions and deletions may occur, Sankowski's
randomized algorithms with one-sided error~\cite{Sankowski04} for transitive
closure imply a worst-case per-update running time of $\bigO(n^{1.575})$ for
the fully dynamic single-source reachability problem.

On the negative side, Henzinger~\ea{}~\cite{HKNS15} showed that unless the Online
Matrix-Vector Multiplication problem can be solved in time
$\bigO(n^{3-\varepsilon})$, $\varepsilon > 0$, no algorithm for the fully dynamic
single-source reachability problem exists with a worst-case update time of
$\bigO(n^{1-\delta})$ and a worst-case query time of $\bigO(n^{2-\delta})$,
$\delta > 0$.
Furthermore, if there is a combinatorial, fully dynamic s-t reachability
algorithm with a worst-case running time of $\bigO(n^{2-\delta})$ per update or
query, then there are also faster combinatorial algorithms for Boolean matrix
multiplication and other problems, as shown by Abboud and Vassilevska
Williams~\cite{AW14}
and Williams and Vassilevska Williams~\cite{WW10}, respectively.

In extensive studies, Frigioni~\ea{}~\cite{FMNZ01} as well as Krommidas and
Zaroliagis~\cite{KZ08} have evaluated a huge set of algorithms for the more
general fully dynamic all-pairs reachability problem experimentally on random
dynamic graphs of size up to \num{700} vertices as well as two static
real-world graphs with randomly generated update operations.
They concluded that, despite their simple-mindedness, static breadth-first or
depth-first search outperform their dynamic competitors on a large number of
instances.
There has also been recent development in designing algorithms that maintain a
reachability index in the static
setting~\cite{ms2014preach,yaiy2013ppl,chwf2013tf,yc2012grail}, which were
evaluated experimentally~\cite{ms2014preach} on acyclic
random and real-world graphs of similar sizes as in this paper.
\section{Algorithms}%
\SectLabel{algorithms}%
We implemented and tested a variety of deterministic, combinatorial algorithms.
An overview is given in \Table{algorithms-overview}.
Additionally, \Table{algorithms} subsumes the corresponding theoretical
worst-case running times and space requirements.
Not all of them are fully dynamic or even dynamic in their original form and
have therefore been ``dynamized'' by us in a more or less straightforward manner.
In this section, we provide a short description of these algorithms, their
implementation, and the variants we considered.
Each algorithm consists of up to four subroutines:
\Routine{initialize()}, %
\Routine{edgeInserted($\Arc{u}{v}$)}, %
\Routine{edgeDeleted($\Arc{u}{v}$)}, %
and \Routine{query($t$)}, %
which define the algorithm's behavior during its initialization phase, in case
that an edge~$\Arc{u}{v}$ is added or removed, and if it is queried whether a
vertex $t$ is reachable from the source, respectively.
We distinguish three groups:
The first group comprises algorithms that are based on static breadth-first and
depth-first search with some improvements.
Algorithms in the second group are based on a simple incremental algorithm that
maintains an arbitrary, not necessarily height-minimal, reachability tree,
and algorithms in the third group use Even-Shiloach trees and thus maintain a
(height-minimal) breadth-first search tree.
We did not implement (and extend to being fully dynamic) the more sophisticated
deletions-only single-source reachability
algorithms~\cite{HKN14,HKN15,CHILP16,BPW19} as they are very involved and
maintain, \eg, a multi-level hierarchy of graphs and node separators, where
Even-Shiloach trees appear only as sub-datastructures.
Due to the resulting huge constants in worst-case time and space complexities,
we expect them to perform much slower in practice.

In the following, we assume an incidence list representation of the graph, \ie,
each vertex has a list of incoming and outgoing edges.
\subsection{Dynamized Static Algorithms}%
Depth-first search (\emph{DFS}) and breadth-first search (\emph{BFS}) are the two classic
approaches to obtain reachability information in a static setting.
Despite their simplicity, studies for \APRlong{}~\cite{FMNZ01,KZ08} report
even their pure versions to be at least competitive with genuine dynamic
algorithms and even superior on various instances.
We consider three variants each:
For our variants \SDFS{} and \SBFS{}
(\AlgLongName{Static DFS/BFS}), we do not maintain any information and start the
pure, static algorithm  for each query anew from the source.
Thus, all work is done in \Routine{query($\cdot$)}.

Second, we introduce a cache as a simple means to speedup queries for our
variants \CDFS{} and \CBFS{} (\AlgLongName{Caching DFS/BFS}).
The cache contains reachability information for \emph{all} vertices and is recomputed
entirely in \Routine{query($\cdot$)} if it has been invalidated by an update.
The rules for cache invalidation are as follows:
An edge insertion is considered \emph{critical} if it connects
a reachable vertex to a previously unreachable vertex.
Similarly, an edge deletion is \emph{critical} if its head is \Reachable{}.
The algorithms keep track of whether a critical insertion or deletion has
occurred since the last recomputation.
The cache is invalidated if either a critical insertion has occurred and
the cached reachability state of a queried vertex $t$ is \Unreachable{}, or
if a critical deletion has occurred and the cached reachability state of $t$ is
\Reachable{}.
Both algorithms may use \Routine{initialize()} to build their
cache.

Finally, we also implemented lazy, caching variants \LDFS{} and
\LBFS{} (\AlgLongName{Lazy DFS/BFS}).
In contrast to the former two, these algorithms only keep reachability
information of vertices they have encountered while answering a query.
As a vertex can only be assumed to be \Unreachable{} if the graph
traversal has been exhaustive, the algorithms each additionally maintain
a flag \Code{exhausted}.
For \Routine{query($t$)}, the cached state of $t$ is hence returned if $t$'s
cached state is \Reachable{} and no critical edge deletion has occurred.
Otherwise, in case that there was no critical edge insertion and $v$'s cached
state is \Unreachable{}, the algorithm has to check the flag
\Code{exhausted}.
If it is not set, the graph traversal that has been started at a previous query
is resumed, thereby updating the cache, until either $t$ is encountered or all
reachable vertices have been visited.
Then, the algorithm returns $t$'s (cached) state.
In all other cases, the cache is invalidated and the traversal must be started
anew.
\begin{table*}[tb]
\caption{Worst-case running times and space requirements.}
\TabLabel{algorithms}
\centering
\astr{1.2}
\begin{tabular}{@{}l@{\hspace{3em}}ccc@{\hspace{3em}}cc@{}}
\toprule
&
&
Time
&
&
\multicolumn{2}{c}{Space}
\\
Algorithm & Insertion & Deletion & Query & Permanent & Update\\
\midrule
\SBFS{},
\SDFS{}
&
$0$
&
$0$
&
$\bigO(n+m)$
&
$0$
&
$\bigO(n)$
\\
\CBFS{},
\CDFS{},
\LBFS{},
\LDFS{}
&
$\bigO(1)$
&
$\bigO(1)$
&
$\bigO(n+m)$
&
$\bigO(n)$
&
$\bigO(n)$
\\
\makecell[cl]{%
\SIMp{\FlagRQ{},\FlagSFQ{},\RestartRatio{}}\\
$\llcorner \qquad \RestartRatio = 0$
}
&
$\bigO(n + m)$
&
\makecell[cc]{%
$\bigO(n \cdot m)$\\
$\bigO(n + m)$\\
}
&
$\bigO(1)$
&
$\bigO(n)$
&
$\bigO(n)$
\\
\makecell[cl]{%
\ESp{\RestartLimit{},\RestartRatio{}},
\MESp{\RestartLimit{},\RestartRatio{}}\\
$\llcorner \qquad \RestartLimit \in \bigO(1) \vee \RestartRatio = 0$
}
&
$\bigO(n + m)$
&
\makecell[cc]{%
$\bigO(n \cdot m)$\\
$\bigO(n + m)$
}
&
$\bigO(1)$
&
$\bigO(n + m)$
&
$\bigO(n)$
\\
\makecell[cl]{%
\SESp{\RestartLimit{},\RestartRatio{}}\\
$\llcorner \qquad \RestartLimit \in \bigO(1) \vee \RestartRatio = 0$
}
&
$\bigO(n + m)$
&
\makecell[cc]{%
$\bigO(n \cdot m)$\\
$\bigO(n + m)$
}
&
$\bigO(1)$
&
$\bigO(n)$
&
$\bigO(n)$
\\
\bottomrule
\end{tabular}
\end{table*}

\subsection{Reachability-Tree Algorithms}%
In a pure incremental setting, \ie, without edge deletions, an algorithm that
behaves like \LDFS{} or \LBFS{}, but updates its cache on edge insertions
rather than queries, can answer queries in $\bigO(1)$ time and spends only
$\bigO(n + m)$ in total for all edge insertions, \ie, its amortized time for an
edge insertion is $\bigO(1)$.
We refer to this algorithm as \SIM{} (\AlgLongName{Simple Incremental}) and describe various
options to make it fully dynamic.
For every vertex $v \in \VertexSet$, \SIM{} maintains a flag
$\Reachable[v]$, which is used to implement \Routine{query($v$)} in
constant time, as well as a pointer $\Code{treeEdge}[v]$ to
the edge in the reachability tree whose head is $v$.
More specifically, the algorithm implements the different operations as
follows:

\noindent
\Routine{initialize()}:
The algorithm traverses the graph using BFS
starting from $\Source$ and sets $\Reachable[v]$ and
$\Code{treeEdge}[v]$ for each vertex $v \in \VertexSet$ accordingly.

\noindent
\Routine{edgeInserted($\Arc{u}{v}$)}: %
If $u$, but not $v$ was reachable before, update \Reachable{} and
$\Code{treeEdge}$ of all vertices that can be reached from $v$ and were
unreachable before by performing a BFS starting at $v$.

\noindent
\Routine{edgeDeleted($\Arc{u}{v}$)}: %
If $\Code{treeEdge}[v] = \Arc{u}{v}$, the deletion of $\Arc{u}{v}$ requires to check
and update all vertices in the subtree rooted at $v$.
We consider two basic options:
Updating the stored reachability information or recomputing it entirely from
scratch.
For the former, we first identify a list \List{} of vertices whose reachability
is possibly affected by the edge deletion, which comprises all vertices
in the subtree rooted at $v$ and is obtained by a simple preorder traversal.
Their state is temporarily set to \Unknown{} and their \Code{treeEdge} pointers are
reset.
Then, the reachability of every vertex $w$ in \List{} is recomputed by
traversing the graph by a backwards BFS starting from $w$ until a reachable
ancestor $x$ is found or the graph is exhausted.
If $w$ is \Reachable{}, the vertices on the path from $x$ to $w$ are added to the
reachability tree using the path's edges as tree edges.
If $w$ is \Unreachable{}, so must be all vertices encountered during the backwards
traversal.
In both cases, this may, thus, reduce the number of vertices with state
\Unknown{}.
Optionally, if $w$ is \Reachable{}, the algorithm may additionally start a forward
BFS traversal from $w$ to update the reachability information of all vertices with
status \Unknown{} in \List{}  that are reachable from $w$.
Moreover, \List{} can be processed
in order either as constructed or reversed.
Independently of this choice, the worst-case running time is in
$\bigO(\Card{\List{}}\cdot(\Card{\List{}} + m))$, as vertices in \List{} may be
traversed $\bigO(\Card{\List{}})$ times by the backwards BFS.
Recomputing from scratch, the second option, requires $\bigO(n + m)$ worst-case
update time.

Thus, our implementation of \SIM{} takes three parameters:
two boolean flags \FlagR{} (negated: \FlagNR{})
and \FlagSF{} (negated: \FlagNSF{}), specifying whether \List{} should be
processed in reverse order and whether a forward search should be started for
each re-reachable vertex, respectively, as well as a ratio $\RestartRatio \in [0,1]$
indicating that if $\List{}$ contains more than $\RestartRatio \cdot n$ elements,
the reachability information for \emph{all} vertices is recomputed from scratch.
\subsection{Shortest-Path-Tree Algorithms}%
In 1981, Even and Shiloach~\cite{ES81} described a simple decremental
connectivity algorithm for undirected graphs that is based on the maintenance
of a BFS tree and requires $\bigO(n)$ amortized update time.
Such a tree is also called Even-Shiloach tree or ES tree for short.
Henzinger~and~King~\cite{HK95} were the first to observe that ES trees
immediately also yield a decremental algorithm for \SSR{} on directed graphs
with the same amortized update time if the source %
is used as the tree's root.
We extend this data structure to make it fully dynamic and consider various
variants.

For every vertex $v \in \VertexSet$, an ES tree maintains its BFS level
$\Level{v}$, which corresponds to $v$'s distance from \Source{}, as well as
an ordered list of in-edges $\InEdgesList{}[v]$.
To efficiently manage this list in the fully dynamic setting, the algorithm
additionally uses an index of size $\bigO(m)$ that maps each edge $\Arc{u}{v}$
to its position in $\InEdgesList{}[v]$.
If $v$ is reachable, its \emph{tree edge} in the BFS tree is the edge with tail
at level $\Level{v} - 1$ whose index $i$ is the smallest in $\InEdgesList{}[v]$
(invariant).
The algorithm stores the index of the tree edge in $\InEdgesList{}[v]$ as
$\TreeEdgeIndex{v}$.
If $v$ is unreachable, $\Level{v} = \infty$ (invariant).
A reachability query \Routine{query($t$)} can thus be answered in $\bigO(1)$ by
testing whether $\Level{t} \neq \infty$.

\noindent
\Routine{initialize()}:
The ES tree is built by a BFS traversal starting from the source.
In doing so, $\InEdgesList{}[v]$ is populated for each vertex $v$ in the order
in which the edges are encountered.
Thus, after the initialization, $\TreeEdgeIndex{v} = 0$.
The update operations are implemented as follows.

\noindent
\Routine{edgeInserted($\Arc{u}{v}$)}: %
Update the data structure in worst-case $\bigO(n+m)$ time by starting a
BFS from $v$ and checking for each vertex that is encountered
whether either its level or, subordinately, its parent index can be decreased.

\noindent
\Routine{edgeDeleted($\Arc{u}{v}$)}: %
If $\Arc{u}{v}$ is $v$'s tree edge, the algorithm tries to find a substitute
edge.
To this end, $v$ is added to an initially empty FIFO-queue \Queue{} containing
vertices whose tree edge and, if necessary, whose level has to be newly
determined.
Vertices in \Queue{} are processed one-by-one as follows:
For each vertex $w$, the index $\TreeEdgeIndex{w}$ is increased until it either
points to an edge with tail at level $\Level{w} - 1$ or $\InEdgesList{}[w]$ is
exhausted.
In the latter case, if $\Level{w} + 1 < n$, $w$'s level is increased by one,
$\TreeEdgeIndex{w}$ is reset to zero, and all children of $w$ in the BFS tree as well
as $w$ itself are added to \Queue{}.
Otherwise, $w$ is unreachable and $\Level{w} := \infty$.
As processing $w$ incurs a runtime cost of $\bigO(\InDegree{w} + \OutDegree{w})$
per level and the number of levels is in $\bigO(n)$, this operation has a
total worst-case running time of $\bigO(n \cdot m)$.

In view of this large update cost, we again introduce an option to alternatively
recompute the BFS tree from scratch.
We use two parameters to control the algorithm's behavior:
a factor $\RestartRatio$ that limits the number of vertices that may be
processed in the queue to $\RestartRatio \cdot n$ as well as an upper bound
\RestartLimit{} on how often a vertex may be (re-)inserted into the queue
before the update operation is aborted and a recomputation is triggered.
We refer to this algorithm as \ES{} (\AlgLongName{Even-Shiloach}).
Observe that if the algorithm recomputes immediately, \ie, if $\RestartRatio =
0$, or each vertex may be processed in \Queue{} only a constant number of times
(and is therefore considered only on a constant number of levels),
\ie, if $\RestartLimit \in \bigO(1)$, the worst-case theoretical running time
is only $\bigO(n + m)$.

We also implemented a variation of \ES{} that sets the tree edge of a vertex
$w$ in the queue directly to the first edge in $\InEdgesList{}[w]$ whose
tail has the lowest level and updates $\Level{w}$ accordingly, which avoids
the immediate re-insertion of $w$ into the queue.
More precisely, while iterating through $\InEdgesList{}[w]$, as realized by
increasing $\TreeEdgeIndex{w}$, this variation keeps track of the minimum level
$l_{\min}$ and the corresponding index $e_{\min}$ of an edge's tail encountered
thereby.
If $\TreeEdgeIndex{w}$ reaches $\Card{\InEdgesList{}[w]}$, \ie, no incoming
edge with tail at level $\Level{w} - 1$ has been found, $\TreeEdgeIndex{w}$ is
set to $0$ and the search continues until $\TreeEdgeIndex{w}$ attains the value
it had when removed from \Queue{}.
Then, $\Level{w}$ is set to $l_{\min} + 1$, $\TreeEdgeIndex{w} = e_{\min}$,
and, if $\Level{w}$ has increased, all children of $w$ in the BFS tree, but not
$w$ itself, are added to \Queue{}.
In consequence, if $w$'s level remains the same, still only a part of
$\InEdgesList{}[w]$ where the index is greater than $\TreeEdgeIndex{w}$ is
scanned.
If $w$'s level has to be increased, however, $\InEdgesList{}[w]$ is scanned
exactly once to determine the new level and tree edge, which is in sharp
contrast to the standard \ES{} algorithm, where $\InEdgesList{}[w]$ is scanned
entirely on each level between the old and the new one, where finally an
incoming edge that is suitable as tree edge is found.
As vertices may skip several levels in one step here, we refer to
this version of \ES{} as \MES{} (\AlgLongName{Multi-Level Even-Shiloach}).

We also consider an even further simplification of \ES{}, \SES{}
(\AlgLongName{Simplified Even-Shiloach}), which does no longer maintain an
ordered list of in-edges for each vertex $v$ and hence also no index
$\TreeEdgeIndex{v}$.
Instead, it stores for each reachable vertex a direct pointer to its tree edge in
the BFS tree.
For each vertex $w$ in \Queue{}, \SES{} simply iterates over all in-edges
in arbitrary order and sets $w$'s tree edge to one whose tail has minimum level.
If this increases $w$'s level, all children of $w$ in the BFS tree are added to
\Queue{}.
Both \MES{} and \SES{} take the same two parameters as \ES{} to control when to
recompute the data structure from scratch.
\section{Experiments}%
\label{sect:experiments}%
\subsection{Environmental Conditions and Methodology}%
We evaluated the performance of all algorithms described in
\Section{algorithms} with all available parameters on both
generated and real-world instances.
All algorithms were implemented in C++17 as part of the open-source algorithms
library \ALGORA{}\cite{Algora}
and compiled with GCC 7 using
full optimization (\texttt{-O3 -march=native -mtune=native}).
Experiments were run on a machine with two Intel Xeon E5-2643 v4 processors
clocked at 3.4 GHz and 1.5TB of RAM under Ubuntu Linux 18.04 LTS with kernel
4.15.
Each experiment was assigned exclusively to one core and bound to allocate
memory only locally.
Instances (except random ones) and algorithm implementations are available
on a dedicated website\footnote{\url{https://dyreach.taa.univie.ac.at}}.

For each algorithm and graph, we measured the time spent during initialization
as well as for each insertion, deletion, and query.
From these, we obtained the total insertion time, total deletion time, total
update time, and total query time as the respective sums.
For the smaller random instances, we ran each experiment three times and use
the medians of the aggregations for the evaluation to counteract artifacts of
measurement and accuracy.

In the following, we use $\Kilo{}$ and $\Mega{}$ as abbreviations
for $\times 10^3$ and $\times 10^6$, respectively.

\subsection{Instances}%
\begin{table*}[h]
\caption{%
Number of vertices $n$,
initial, average, and final number of edges
$\initE$, $\avgE$, and $\finE$,
average density $\avgD$,
total number of updates $\Updates$,
and percentage of additions $\AddsRatio$
among updates, %
of real-world instances.}
\TabLabel{realworld-instances}
\centering
\astr{1.2}
\begin{tabular}{@{}lrrrrrrrrr@{}}
\toprule
Instance & $n$ & $\initE$ & $\avgE$ & $\finE$ & $\avgD$ & $\Updates$ & $\AddsRatio$ \\
\midrule
\texttt{FR}           & \num{2.2}\Mega &     \num{3} & \num{13.0}\Mega & \num{24.5}\Mega &  \num{5.9} & \num{59.0}\Mega{} & \SI{71}{\percent} \\
\texttt{DE}           & \num{2.2}\Mega &     \num{4} & \num{16.7}\Mega & \num{31.3}\Mega &  \num{7.7} & \num{86.2}\Mega & \SI{68}{\percent}  \\
\texttt{IT}           & \num{1.2}\Mega &     \num{1} &  \num{9.3}\Mega & \num{17.1}\Mega &  \num{7.8} & \num{34.8}\Mega & \SI{75}{\percent}  \\
\texttt{NL}           & \num{1.0}\Mega &     \num{1} &  \num{5.7}\Mega & \num{10.6}\Mega &  \num{5.4} & \num{20.1}\Mega & \SI{76}{\percent}  \\
\texttt{PL}           & \num{1.0}\Mega &     \num{1} &  \num{6.6}\Mega & \num{12.6}\Mega &  \num{6.4} & \num{25.0}\Mega & \SI{75}{\percent}  \\
\texttt{SIM}       &  \num{100}\Kilo &     \num{2} &   \num{401}\Kilo &   \num{747}\Kilo &  \num{5.9} &  \num{1.6}\Mega & \SI{73}{\percent}  \\
\midrule
\texttt{AS-CAIDA}     &   \num{31}\Kilo & \num{73}\Kilo & \num{99.9}\Kilo & \num{113}\Kilo & \num{4.7} & \num{1.4}\Mega & \SI{51}{\percent} \\
\midrule
\texttt{FR\_SHUF}     & \num{2.2}\Mega &   \num{4.0} & \num{16.4}\Mega & \num{30.4}\Mega & \num{7.4} & \num{53.1}\Mega & \SI{79}{\percent} \\
\texttt{DE\_SHUF}     & \num{2.2}\Mega &   \num{3.8} & \num{22.6}\Mega & \num{41.1}\Mega & \num{10.4} & \num{76.4}\Mega & \SI{77}{\percent} \\
\texttt{IT\_SHUF}     & \num{1.2}\Mega &   \num{3.8} & \num{10.9}\Mega & \num{20.5}\Mega & \num{9.1} & \num{31.4}\Mega & \SI{83}{\percent} \\
\texttt{NL\_SHUF}     & \num{1.0}\Mega &   \num{3.8} & \num{6.7}\Mega  & \num{12.6}\Mega  & \num{6.4} & \num{18.1}\Mega & \SI{85}{\percent} \\
\texttt{PL\_SHUF}     & \num{1.0}\Mega  &  \num{3.6} & \num{7.9}\Mega & \num{14.9}\Mega & \num{7.7} & \num{22.7}\Mega & \SI{83}{\percent} \\
\texttt{SIM\_SHUF}    &  \num{100}\Kilo & \num{5.6} & \num{476}\Kilo & \num{892}\Kilo & \num{4.7} & \num{1.6}\Mega & \SI{80}{\percent} \\
\bottomrule
\end{tabular}
\end{table*}

\paragraph*{Random Instances.}
To assess the average performance of our algorithms,
we generated a set of smaller random directed graphs according to the
Erd\H{o}s-Renyí model $G(n, m)$ with $n = \num{100}\Kilo{}$ vertices and $m = d
\cdot n$ edges, where $d \in [ \num{1.25} \dots \num{50}]$, in each case along
with a random sequence of operations $\Ops$ consisting of edge insertions, edge
deletions, as well as reachability queries.
In the same fashion, we generated a set of larger instances with $n =
\num{10}\Mega$ vertices and $m = d \cdot n$ edges.
For insertions, we drew pairs of vertices uniformly at random from
$\VertexSet$, allowing also for parallel edges.
For deletions and reachability queries, each edge or vertex, respectively, was
equally likely to be chosen.
For a fixed source vertex, we tested sequences of $\sigma=\num{100}\Kilo{}$
operations, where insertions, deletions, and queries appear in batches of
ten, but are processed individually by the algorithms.
We evaluated different proportions of the three types of operations.

\paragraph*{Kronecker Instances.}
Reachability plays an important role in the analysis of social networks, whose
structures differ greatly from that of Erd\H{o}s-Renyí graphs,
\eg, in terms of degree distribution.
Our test instances therefore additionally include stochastic Kronecker
instances~\cite{Leskovec2010}, which were shown to model the structure of such
networks very well. %
We generated two sets containing \num{20} instances each, using the
\texttt{krongen} tool that is part of the \SNAP{} software library
\cite{leskovec2016snap} and the estimated initiator matrices given
in~\cite{Leskovec2010} that correspond to real-world networks.
To obtain dynamic graphs, we generated a sequence of different \emph{snapshot
graphs} for each initiator matrix, computed the differences between two
subsequent instances, and simulated an update sequence by applying them in
random order.
In the first set, we used sequences of ten graphs that were generated in
\num{17} iterations with up to $\approx$\num{130}\Kilo{} vertices each, whereas
in the second, the graphs in each sequence were generated with increasing
number of iterations, starting from five up to \num{17}, which resulted in
instances having around \num{30} vertices initially and again up to
$\approx$\num{130}\Kilo{} in the end.
We refer to the first set as \texttt{kronecker-csize} and to the second as
\texttt{kronecker-growing}.
For each dynamic graph, we used the ten vertices with highest out-degree in
their respective initial graph as sources.
All instances in \texttt{kronecker-csize} have densities between \num{0.7} and
\num{16.4}.
Their update sequences consist of equally many insertions and deletions, whose
lengths range between \num{1.6}\Mega{} and \num{702}\Mega{}.
In \texttt{kronecker-growing}, the densities vary between \num{0.9} and
\num{16.4}.
There are \num{282}\Kilo{} to \num{82}\Mega{} updates, \SI{66}{\percent} to
\SI{75}{\percent} of which are insertions.
\paragraph*{Real-World Instances from \KONECT{}.}
Our set of test instances is complemented by a collection of real-world dynamic
networks, which also includes real-world update sequences.
For algorithms that maintain a reachability tree, the latter is especially of
interest, as the selection and order of edge insertions and deletions may
affect the amount of work required immensely.
We used all six directed, dynamic instances available from the Koblenz Network
Collection \KONECT{}~\cite{konect}, %
a collection of real-world graphs from various application scenarios.
The graphs are given as a list of edge insertions and deletions, each of which
is assigned a timestamp, and model the hyperlink structure between Wikipedia
articles for six different languages.
Hyperlink networks are a variant of social networks,
where reachability information is used, \eg, to detect dependencies or
topical clusters.
For our evaluation, the edge insertions and deletions with the smallest
timestamp form the initial graph, and all further updates are grouped by their
timestamp.
We set the source vertex to be the tail of the first edge with minimum timestamp.
Our instances have between \num{100}\Kilo{} (simple English) and
\num{2.2}\Mega{} vertices (French) and from initially less than five up to
\num{747}\Kilo{} to \num{24.5}\Mega{} edges, which result from between
\num{1.6}\Mega{} and
\num{86}\Mega{} update operations, consisting of both edge insertions and
deletions.
We refer to these instances as \texttt{FR},
\texttt{DE}, \texttt{IT}, \texttt{NL}, \texttt{PL},
and \texttt{SIM}.

To see whether differences in the algorithms' performance are rather due to the
structure of the graphs or the order of updates, we generated five new,
``shuffled'' instances per language by randomly assigning new timestamps to the
update operations, which we refer to as \emph{shuffled \KONECT{}}.
As for the original instances provided by \KONECT{}, we ignored removals of
non-existing edges.

\paragraph*{Real-World Instances from \SNAP{}.}
Additionally, we use a collection of \num{122} snapshots of the computer network
describing relationships in the CAIDA Internet Autonomous System, which is made
available via the Stanford Large Network Dataset Collection
\SNAP{}~\cite{snap}.
We built a dynamic, directed graph \texttt{AS-CAIDA} with $n=\num{31}\Kilo{}$
and $m=\num{73}\Kilo{}$ to~$\num{113}\Kilo{}$ from this collection by
using the differences between two subsequent snapshots as updates.
Edges are directed from provider to customer and there is a pair of
anti-parallel edges between peers and siblings.
We obtained ten instances from this graph by choosing one of the ten vertices
with highest out-degree, respectively, as source.

\Table{realworld-instances} lists the detailed numbers for all real-world
instances and the respective average values for the shuffled \KONECT{}
instances.
In each case, the updates are dominated by insertions, which constitute
\SI{51}{\percent} for \texttt{AS-CAIDA}, \SIrange{68}{76}{\percent} for
\KONECT{},
and \SIrange{77}{85}{\percent} for shuffled \KONECT{}.
The average density varies between \num{4.7} (\texttt{AS-CAIDA}, \texttt{SIM\_SHUF}) and
\num{10.4} (\texttt{DE\_SHUF}).

\input{figures_plot-random100k}
\subsection{Experimental Results}%
\subsubsection{Random graphs}%
For $n = \num{100}$\Kilo{}, we generated \num{20} graphs per density $d =
\frac{m}{n}$ along with a sequence of \num{100}\Kilo{} operations, where edge
insertions, edge deletions, and queries were equally likely.
In consequence, the density of each dynamic graph remains more or less constant
during the update sequence.
The timeout was set to one hour.
\Figure{random-100k} depicts the results, which we will discuss in the
following.
Note that, except for \Figure{random-100k-all-init-abs}, the plots use
logarithmic axis in both dimensions.

\paragraph*{Relative Performances within Groups
(\FigureSubRange{random-100k-static-queries-rel}{random-100k-xes-deletions-rel}).}%
For the discussion of the results, we group the algorithms as in
\Section{algorithms}.
The \textbf{first group} consists of the six \textbf{dynamized static algorithms}
\SBFS{}, \SDFS{}, \CBFS{}, \CDFS{}, \LBFS{}, and \LDFS{}.
Recall that all work is done in \Routine{query($\cdot$)} here,
which is why we evaluate them based on their mean total query time.
\Figure{random-100k-static-queries-rel} shows the relative performance of this
algorithm group compared to \LBFS{}, which was the \emph{best algorithm} on
average over all densities and for each density always seven to \num{16} times
faster on average than the ``pure'' static algorithms \SBFS{} and \SDFS{}.
Up to a density of \num{4.5}, \LBFS{} is beaten by \LDFS{}, however, the
performance gap between \LBFS{} and \LDFS{} increases at least linearly as the
graphs become denser.
The eager caching versions \CBFS{} and \CDFS{} show similar performance to
their lazy counterparts on sparse graphs, but then deteriorate exponentially
compared to the latter and eventually even fall behind the pure static
variants \SBFS{} and \SDFS{}, respectively.
\emph{To summarize, the algorithms based on DFS are only faster than
their BFS-based counterparts on sparser instances and
distinctly slower on denser ones.}

The \textbf{second group} of algorithms consists of the fully dynamic variants
of the \textbf{simple incremental} algorithm \SIM{}.
These algorithms only differ in their implementation of
\Routine{edgeDeleted($\cdot$)} and, thus, we evaluate them on their mean
deletion time.
We tested different combinations of the boolean flags \FlagR{} and \FlagSF{}
along with different values for the recomputation threshold~$\RestartRatio$.
\emph{One main observation is that, regardless of $\RestartRatio$, the algorithms
\SIMp{\FlagNR{}/\FlagSF{}/\RestartRatio{}} were faster than the algorithms using
other combinations of the flags, but the same value \RestartRatio{}},
where the worst-performing was \SIMp{\FlagNR{}/\FlagNSF{}/\RestartRatio}.
If the flags \FlagRQ{} and \FlagSFQ{} were fixed,
smaller values for \RestartRatio{} showed better performance than larger,
except for extremely small ones.
Recall that if \RestartRatio{} is zero, the algorithm always discards its
current reachability tree and recomputes it from scratch using
BFS, whereas if \RestartRatio{} is one, it always reconstructs a
reachability tree.
Hence, \RestartRatio{} may be seen as a means to control outliers that
necessitate the re-evaluation of the reachability of a large number of
vertices.
To keep the number of variants manageable,
\Figure{random-100k-si-deletions-rel} only shows the
relative mean total deletion time of
\SIM{} with four different parameter sets:
\FlagNR{}/\FlagSF{} with $\RestartRatio{}=\num{0.25}$, $\RestartRatio{}=\num{0.5}$, and
$\RestartRatio{}=\num{1}$, respectively, and \FlagNR{}/\FlagNSF{} with
$\RestartRatio{}=\num{0.25}$.
The \emph{fastest algorithm} on average across all densities in this set was
\SIMp{\FlagNR{}/\FlagSF{}/.25}, which is therefore also used as reference.
The same algorithm with disabled forward search, \ie,
\SIMp{\FlagNR{}/\FlagNSF{}/.25}, was up to a factor of around \num{16} slower
on sparse graphs.
As the graphs become denser, this factor decreases exponentially down to
less than \num{1.5} for graphs having $d=\num{40.0}$ and above.
The reason for this will be discussed with
\Figure{random-100k-fast-deletions-abs}.
\SIMp{\FlagNR{}/\FlagSF{}/.5} and \SIMp{\FlagNR{}/\FlagSF{}/1}
show similar performance as \SIMp{\FlagNR{}/\FlagSF{}/.25} for densities
of at least \num{1.5} and \num{2.0}, respectively,
however with extreme spikes at $d=\num{2.5}$
and $d=\num{4.0}$ if $\RestartRatio = \num{1}$, which are caused by few
instances with enormous cost for re-establishing the reachability tree.
For $d=\num{2.5}$, \eg, \SIMp{\FlagNR{}/\FlagSF{}/1} needed around
\SI{10}{\minute} for one specific edge deletion operation on one graph, whereas
the maximum deletion time on all other instances was less than
\SI{150}{\milli\second}.
The total deletion time hence was less than \SI{1}{\second} on \num{19} instances
and around \SI{10}{\minute} on the \num{20}th, which resulted in a mean
total deletion time of \SI{33.7}{\second} for \SIMp{\FlagNR{}/\FlagSF{}/1} on
graphs with density \num{2.5}.
By contrast, the mean total deletion time of \SIMp{\FlagNR{}/\FlagSF{}/.25} on these
\num{20} instances was \SI{619}{\milli\second}.
The other spikes can be explained similarly.
\emph{In conclusion, low values for $\RestartRatio$ can effectively control outliers
and speed up the average deletion time by factors of up to $\num{307}$.}

The \textbf{third group} of algorithms comprises those based on \textbf{ES
trees}: \ES{}, \MES{}, and \SES{}.
We tested each of them with different values for the parameters \RestartLimit{}
and \RestartRatio{}.
Here, both parameters serve to limit excessive update costs that occur when
either the levels of a smaller set of vertices in the ES tree
increase multiple times (\RestartLimit{}) or a large set of vertices is
affected (\RestartRatio{}).
They turned out to be very useful.
We tested three parameter sets:
An early abortion of the update process and recomputation with
$\RestartLimit=\num{5}$ and $\RestartRatio = \num{0.5}$,
a late variant with
$\RestartLimit=\num{100}$ and $\RestartRatio = \num{1}$,
and finally
$\RestartLimit=\infty$ and $\RestartRatio = \infty$,
which does not impose any limits.
Similar as in case of \SIM{}, the algorithms only differ in their
implementation of \Routine{edgeDeleted($\cdot$)}. %
\Figure{random-100k-es-deletions-rel} reports the mean total deletion time
relative to the (on average) \emph{best algorithm} in this set, \SESp{5/.5}.
For sparse graphs, the \ES{} algorithms were up to approximately \num{1400}
times slower than \SESp{5/.5}.
This factor drops super-exponentially as the graphs become denser and reaches a
value of around \num{1} near $d=\num{12}$.
The unlimited variants %
showed an even worse performance on graphs up to a density of \num{4.0} with
several timeouts, but a performance similar to, or, in case of \ES{}, an even
better one than their limited versions for denser graphs.
In all cases, the timeouts occurred on six instances with $d=\num{1.5}$, four
with $d=\num{1.75}$, and one with $d=\num{2.5}$ and $d=\num{4.0}$,
respectively.
For $d=\num{1.25}$, \ESp{$\infty$/$\infty$} timed out four times, whereas the
\MES{} and \SES{} variants only once.
Apart from one exception, all timeouts were caused by a single deletion
operation that took more than one hour, in some cases even more than five.

Differences between the limited versions of \MES{} and \SES{} are barely
observable on this scale.
\Figure{random-100k-xes-deletions-rel} zooms in on the values of interest for
these algorithms.
Evidently, \SESp{5/.5} outperformed \MESp{5/.5} both on very sparse instances up
to $d=\num{1.75}$ as well as on denser ones from $d=\num{14}$ and onward.
In the middle range, it was less than \SI{6}{\percent} slower than
\MESp{5/.5}.
Recall that in contrast to \SES{}, \MES{} stores information about
the incoming edges of a vertex.
However, for very sparse as well as denser instances, the additional knowledge
available to \MES{} seemingly cannot outweigh the increased workload that comes
with the maintenance of this information:
In the former case, the list of in-edges is short and therefore scanned very
quickly in \SES{}, whereas in the latter, a replacement tree edge with
tail on the same level can be expected to be found very early in \SES{}'s
scanning process.
\emph{To summarize,
for both \SES{} and \MES{}, the variants that are more reluctant to
recompute from scratch performed slightly worse than their respective counterparts.
The \ES{} algorithms were almost always outperformed by %
\MES{} and \SES{}.}

\input{figures_plot-ratios}
\input{figures_plot-ratios-queries-oneline}
\paragraph*{Update Performances (\FigureSubRange{random-100k-fast-additions-rel}{random-100k-fast-updates-rel}).}
Next, we compare the relative performances of the \SIM{} and the
\ES{}/\MES{}/\SES{} algorithm classes using
\SIMp{\FlagNR/\FlagSF/.25},
\SIMp{\FlagNR/\FlagNSF/.25},
\MESp{5/.5}, and
\SESp{5/.5} as representatives.
\Figure{random-100k-fast-additions-rel} depicts the mean average total
insertion times.
Despite identical implementation, \SIMp{\FlagNR/\FlagSF/.25} was slightly
faster than \SIMp{\FlagNR/\FlagNSF/.25} on sparser instances, which may be due
to structural differences in their reachability trees.
\MESp{5/.5} and \SESp{5/.5} were four to approximately \num{16} times slower
than \SIMp{\FlagNR/\FlagSF/.25}, where the maximum was reached at a density of
$\num{5.0}$.
These experimental results conform with the theoretical performance analysis of
\SIM{}, which yields a ``perfect'' amortized update time of $\bigO(1)$ in the
incremental setting.
\MESp{5/.5} is slightly slower than \SESp{5/.5} %
due to the additional information it maintains.
The overall situation is inverted in case of deletions, as
\Figure{random-100k-fast-deletions-rel} shows.
Here, \MESp{5/.5} and \SESp{5/.5} outperformed both \SIMp{\FlagNR/\FlagSF/.25}
and \SIMp{\FlagNR/\FlagNSF/.25}, the latter even by a factor of almost \num{24}
on very sparse instances.
\SIMp{\FlagNR/\FlagSF/.25} was \SI{15}{\percent} to \SI{100}{\percent}
slower on average than \SESp{5/.5}.

These findings suggest that \SIMp{\FlagNR/\FlagSF/.25} would be the best
choice among these algorithms unless the proportion of edge deletions is
markedly high.
However, insertions and deletions are not equally costly,
as \Figures{random-100k-fast-additions-abs}{random-100k-fast-deletions-abs}
demonstrate.
The minimum and maximum total mean running times for deletions were roughly
\num{50} times higher than those for insertions.
Moreover, they show that the effort to process an update decreases at least
exponentially as the density increases.
The reason for this observation is twofold:
First, the probability that a newly inserted edge will be part of the
reachability tree or that an edge of the current reachability tree is deleted
diminishes as the density grows.
This holds especially for the algorithms of the second group, which do not care
about the distance from the source vertex, and where the reduction in the total
insertion time is pronounced.
Second, if a deletion of a tree edge really occurs, a replacement edge is,
intuitively speaking, usually not ``too far'' and the reachability tree
can be mended with little expense.
The latter also speeds up \SIMp{\FlagNR/\FlagNSF/.25}'s process of handling
edge deletions, as the relatively costly (and numerous) backwards
breadth-first searches terminate quickly.
\Figure{random-100k-fast-updates-rel} depicts the relative mean total update
times, where insertions and deletions occur with equal probability.
As deletions are distinctly more time-consuming than insertions%
---\SESp{5/.5} and \MESp{5/.5} spent $\approx$\,\num{70}--\SI{85}{\percent},
\SIMp{\FlagNR/\FlagSF/.25} even $\approx$\,\num{94}--\SI{99}{\percent} of the
update time on deletions---%
\emph{\SESp{5/.5} showed the best performance on average over all densities.}
Again, \MESp{5/.5} was slower on very sparse and slightly denser instances by up to
about \SI{20}{\percent}.
\SIMp{\FlagNR/\FlagSF/.25}'s performance was
roughly similar to \MESp{5/.5}'s, however with a largest deviation of
\SI{58}{\percent} from \SESp{5/.5}'s at $d=\num{40.0}$.

Even though it is of less importance in the case of long operation sequences,
we take a brief look at the initialization time, as shown in
\Figure{random-100k-all-init-abs}.
The algorithms are split into three groups here:
Whereas \SBFS{}, \SDFS{}, \LBFS{}, and \LDFS{} do not use this phase,
all other algorithms traverse the graph once and build up their data structures.
\CBFS{}, \CDFS{}, \SIM{}, and \SES{} reserve and access $\bigO(n)$
space, but \ES{} and \MES{} need to setup $\bigO(n + m)$ space,
which is clearly reflected in the running time, as
\Figure{random-100k-all-init-abs} shows.
Note that \Figure{random-100k-all-init-abs} does \emph{not} use
logarithmic scales.

\paragraph*{Overall Performances.} %
\textbf{\Figures{random-100k-all-total-rel}{random-100k-all-total-abs}} depict
the mean total running time if insertions, deletions, and queries occur with
equal probability.
The fastest dynamized static algorithm, \LBFS{}, was clearly outperformed by
\SIMp{\FlagNR/\FlagSF/.25}, \MESp{5/.5}, and \SESp{5/.5} on all densities.
For sparser graphs up to $d=\num{4.0}$, however,
the lazy and caching variants were faster than \ES{}.
On dense instances, where the update costs decrease rapidly, the initialization
time begins to show through for \SIM{} and the \ES{} family.
\emph{The \SES{} algorithms performed best in these experiments, with \SESp{5/.5}
being the overall fastest on average}.

\input{figures_plot-random10m-all}
\paragraph*{Ratios of Insertions, Deletions, and Queries (\Figure{random-100k-ratios}).}
We next investigate whether and how the picture changes if the proportion of
insertions and deletions varies.
Taking up on the observation that the \SIM{} algorithms were considerably
faster on insertions than \MES{} and \SES{}, but slower on deletions,
we compare the performance of the fastest of each of them, \ie,
\SIMp{\FlagNR/\FlagSF/.25}, \MESp{5/.5}, and \SESp{5/.5} on
random instances with $n = \num{100}\Kilo{}$ vertices,
different initial densities
$d \in \Set{\num{2.5}, \num{5},\num{10}, \num{20}}$,
and $\Ops = \num{100}\Kilo{}$.
We sampled ten graphs per density.
As unequal ratios of insertions and deletions change the density of the graphs
over time, \Figure{random-100k-ratios} shows the mean total update time divided by the
average number of edges.
As expected, \MESp{5/.5}, and \SESp{5/.5} outperformed \SIMp{\FlagNR/\FlagSF/.25}
for low ratios of insertions, whereas the
opposite holds if there are many insertions among the updates.
\emph{The threshold is around \SI{50}{\percent} for all densities.}  \MESp{5/.5} was
similarly fast as \SESp{5/.5} if the proportion of deletions was high (and $d$
is small), and became relatively slower as the ratio of insertions grew.

In our setting, all dynamized static algorithms were clearly inferior.
We expected a performance increase if queries occur either very rarely or, if
a cache is used, very frequently.
We reviewed this assumption experimentally by examining the performance of
\CBFS{}, \CDFS{}, \LBFS{}, and \LDFS{} in comparison to
\SIMp{\FlagNR/\FlagSF/.25}, \MESp{5/.5}, and \SESp{5/.5} for varying ratios of
queries among the operations.
We did not include \SBFS{} and \SDFS{}, as \LBFS{} and \LDFS{} are always at
least as fast.
We again sampled ten instances with $n = \num{100}\Kilo{}$ vertices for each
density $d \in \Set{\num{2.5}, \num{5},\num{10}, \num{20}}$, in each case along
with $\Ops = \num{100}\Kilo{}$ operations.
To keep the density of the graphs constant, insertions and deletions occur with
equal probabilities.
\Figure{random-100k-ratios-queries} depicts the mean total operation times.
\emph{Although the results confirm our assumption, none of the dynamized static
algorithms can compete with the dynamic ones, neither for sparse nor for
denser graphs.}

\input{figures_plot-kronecker}
\input{figures_plot-konect}
\paragraph*{Large Graphs (\Figure{random-10m-all}).}
We repeated our experiments on larger graphs with
$n=\num{10}\Mega{}$ vertices for
the algorithms
\MES{},
\SES{},
and
\SIM{}.
\Figure{random-10m-all} shows the
absolute mean total insertion, deletion, and updates times
as well as the
mean total insertion time relative to
\SIMp{\FlagNR/\FlagSF/.25},
the total deletion time relative to \SESp{5/.5},
and the mean total update time relative to \SESp{5/.5}.
As for the instances with $n=\num{100}\Kilo{}$, the update time was dominated
heavily by the deletion time and decreased with growing density.
The mean total update time relative to \SESp{5/.5} here almost
equals the deletion time.
\SES{} and \MES{} with parameters \texttt{100/1} were almost identical to
their more restricted counterparts and are therefore not shown.
As before, \SIM{} outperformed \MES{} and \SES{} for insertions, %
whereas the latter two were faster than \SIM{} for deletions.
\emph{In both cases, the picture is similar to that for the smaller instances,
however, the speedup factor has increased markedly.}
In total, \SIMp{\FlagNR/\FlagSF/.25} was \num{2.5} to \num{10} times slower than
the \emph{best algorithms} \MES{} and \SES{}, which in turn performed almost
identically.
\subsubsection{Kronecker Instances}%
So far, we only assessed the algorithms' performance on random graphs generated
according to the Erd\H{o}s-Renyí model.
Kronecker instances mimic real-world networks and hence exhibit a different
structure. %
The results for the \texttt{kronecker-csize} graphs are shown in
\Figure{kronecker-csize}.
\emph{On all instances,
\SESp{5/.5} outperformed the other algorithms}, but was closely
followed by \MESp{5/.5}, whereas \SIMp{\FlagNR/\FlagSF/.25} was two to
\num{15.2} times slower and \SIMp{\FlagNR/\FlagNSF/.25} with slowdown factors
of \num{6.9} to \num{57.5} was far from being competitive.
Similar to the random instances, at least \SI{71}{\percent}
of the update time was spent on deletions,
with one exception (\textsf{email-inside}, \SI{51}{\percent}).
Despite their higher insertion rate, the results for
\texttt{kronecker-growing} are similar (cf.~\Figure{kronecker-growing}).
\emph{All in all, the picture is consistent with that on %
random instances.}
\subsubsection{Real-World Graphs}%
We evaluated the algorithms %
\MES{}, \SES{}, and \SIM{} also on real-world graphs that come with real-world
update sequences.
\Figure{konect} shows in the upper part the update times for the original
\KONECT{} instances as well as the mean update time of the \SNAP{} instances.
On all these instances, the algorithms spent more than \SI{89}{\percent} of the
update time on deletions.
Without exception,
\SIMp{\FlagNR/\FlagSF/.25} here distinctly \emph{outperforms}
all competitors, followed with considerable distance by \SESp{5/.5}, which is
in turn faster than \MESp{5/.5} by several factors.
On \texttt{AS-CAIDA}, which has a mostly random update sequence, the lead of
\SIMp{\FlagNR/\FlagSF/.25} is clearly less, but still visible.

The overall picture did not change for the shuffled \KONECT{} instances with
updates in random order, as depicted at the bottom of \Figure{konect}.
However, the speedup of \SIMp{\FlagNR/\FlagSF/.25} in comparison to \SESp{5/.5}
decreased visibly in general, from up to \num{23} to a maximum of less than
seven.
The performance ratio of \MESp{5/.5} and \SESp{5/.5} remained constant.
As the graphs at each point in time can be assumed to have similar
characteristics as the Kronecker instances, these
results demonstrate
that the order of the updates (random or not) influences the
performance of  \SIMp{\FlagNR/\FlagSF/.25}  and \SESp{5/.5}, but it can only
partially explain that the former performs better on real-world graphs than
on Kronecker and random graphs.
Since deletions are significantly slower than insertions, we investigated
the percentage of ``expensive'' deletions, \ie, deletions that change the
reachability tree.
For \SIMp{\FlagNR/\FlagSF/.25}, this number is at most \SI{13}{\percent}
on the \KONECT{} graphs, but up to \SI{33}{\percent} on Kronecker
graphs with comparable density.
For \SESp{5/.5}, however, this number is \SI{19}{\percent} and
\SI{29}{\percent}, respectively.
As the deletion time is much higher for \SIM{} than for \SES{}, this can
explain the difference in performance on real-world vs.\ Kronecker graphs.
A reason for the relatively small change percentage in real-world graphs for
\SIMp{\FlagNR/\FlagSF/.25} might be that the distribution of lifetimes of the
edges is different in real-world and Kronecker graphs.
In the latter, the probability that an edge exists in two subsequent snapshot
graphs is very low, implying that the lifetime of every edge is relatively
small, in contrast to edges representing hyperlinks between articles, as in the
\KONECT{} graphs.
\section{Conclusion}%
The simplified Even-Shiloach algorithm, \SES{}, with parameters \texttt{5/.5}
showed the best performance on all instances except for the real-world dynamic
graphs, where it was outperformed by the fully dynamic
version of the simple incremental algorithm, \SIM{}, with parameters
\texttt{\FlagNR{}/\FlagSF{}/.25}.
However, \SES{} was in particular superior in handling edge deletions, which heavily
dominated the update costs across all tested sets of instances.
All algorithms benefitted considerably from introducing recomputation
thresholds.
Breadth-first search and depth-first search, even with enhancements,
were unable to compete with the dynamic algorithms, irrespective
of the proportion of queries.
The impact of degree distribution on the algorithms' performance remains
unclear.

In a nutshell:
We recommend to use \SIMp{\FlagNR{}/\FlagSF{}/.25} on real-world networks with
long-living edges or if the ratio of insertions is distinctly above
\SI{50}{\percent}, and otherwise \SESp{5/.5}.

\bibliography{paper}
\end{document}